\documentclass[preprint,12pt]{elsarticle} 
\usepackage{wrapfig}
\usepackage{graphicx}
\usepackage{amsmath}

\newcommand{\comment}[1]{}
\journal{Physics and Chemistry of Solids}
\begin{document}
\begin{frontmatter}
\title{Possible Structural Transformation and Enhanced Magnetic Fluctuations in Exfoliated $\alpha$-RuCl$_3$}
\author[WUSTL]{Boyi Zhou}
\author[Boston College]{Yiping Wang}
\author[Boston College]{Gavin B. Osterhoudt}
\author[ORNL]{Paige Kelly}
\author[UT,ORNL]{David Mandrus}
\author[UNI]{Rui He}
\author[Boston College]{Kenneth S. Burch}
\author[WUSTL,IMSE]{Erik Henriksen}
\address[WUSTL]{Department of Physics, Washington University in St.~Louis, 1 Brookings Dr., St.~Louis, MO 63130, USA}
\address[Boston College]{Boston College Department of Physics, 140 Commonwealth Avenue, Chestnut Hill, MA 02467, USA}
\address[ORNL]{Material Science \& Technology Division, Oak Ridge National Laboratory, Oak Ridge, Tennessee 37831, USA}
\address[UT]{7 Department of Material Science and Engineering, University of Tennessee, Knoxville, Tennessee 37996, USA}
\address[UNI]{Department of Electrical and Computer Engineering, Texas Tech University, Lubbock, Texas 79409, USA}
\address[IMSE]{Institute for Materials Science \& Engineering, Washington University in St.~Louis, 1 Brookings Dr., St.~Louis, MO 63130, USA}

\begin{abstract}
We present initial Raman spectroscopy experiments on exfoliated flakes of $\alpha$-RuCl$_3$, from tens of nm thick down to single layers. Besides unexpectedly finding this material to be air stable, in the thinnest layers we observe the appearance with decreasing temperature of a symmetry-forbidden mode in crossed polarization, along with an anomalous broadening of a mode at 164 cm$^{-1}$ that is known to couple to a continuum of magnetic excitations. This may be due to an enhancement of magnetic fluctuations and evidence for a distorted honeycomb lattice in single- and bi-layer samples.
\end{abstract}

\begin{keyword}
A. magnetic materials
C. Raman spectroscopy
D. phonons
\end{keyword}
\end{frontmatter}
\section{Introduction}
\begin{figure}
\vspace{-6ex}
\begin{center}
\includegraphics[width=0.4\textwidth]{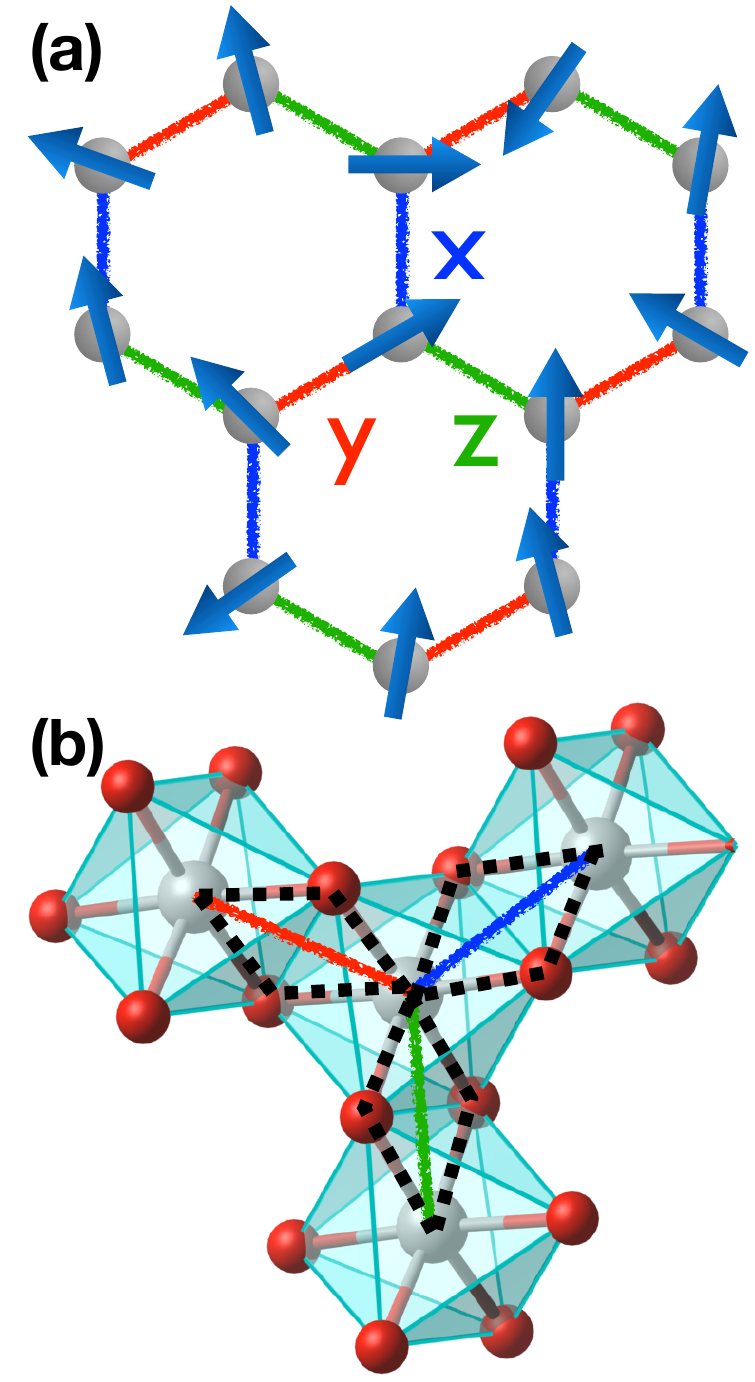}
\end{center}
\vspace{-4ex}
\caption{\label{fig:Kit} (a) Kitaev model of spin 1/2 on the honeycomb lattice with the three in-equivalent bonds indicated by the interacting component of the spin. (b) Structure of the ideal $\alpha$-RuCl$_{3}$ lattice, with Cl atoms octahedrally coordinated around each Ru. The hopping pathways are indicated by black dashed lines, which for a perfect $J_{eff}=1/2$ state will lead to a Kitaev interaction.}
\vspace{-4ex}
\end{figure}

Just over a decade ago, Kitaev proposed a model of a many-body ground state that can be exactly solved, yielding a spin liquid with Majorana Fermion excitations~\cite{Kitaev:2006ik,Nasu:2016hga}. He considered a honeycomb lattice with spin 1/2 moments experiencing a bond-dependent exchange, such that different components of the spin couple along different bonds (see Figure~\ref{fig:Kit}a). This model can be realized in materials under the right conditions of crystal electric field, spin-orbit coupling and on-site Coulomb repulsion that produce an insulator with $J_{eff}=1/2$ moments. In systems where the honeycomb lattice is formed by placing the magnetic atom inside edge-sharing octahedra, one can realize the necessary bond-dependent exchange due to the impact of strong spin-orbit coupling on the hopping (see Figure~\ref{fig:Kit})~\cite{Jackeli:2009hz,Kim:2008gi,2017arXiv170107837C}. A key difficulty with this proposal is that additional interaction terms may arise and produce long range order~\cite{Kim:2015iq,Yadav:2016br,Mazin:2012gq,Gretarsson:2013fp}. While some of these terms are enabled simply by symmetry, they are strongly enhanced by lattice distortions that mix the $J_{eff}=1/2$ and $J_{eff}=3/2$ states, altering the hopping terms. Recently, $\alpha$-RuCl$_3$ has emerged as a potential candidate to realize a Kitaev quantum spin liquid state~\cite{Plumb:2014hh,Sandilands:2015hn,Nasu:2016hga,Johnson:2015jc,Banerjee:2016jz,Sandilands:2016gr,Cao:2016ep,Kubota:2015gu,Lang:2016jn,Zhou:2016fe,Ran:2017ke,Banerjee:2017dk}. 

IR, Raman and photo-emission spectroscopy combined with DFT calculations strongly suggest the system is close to the $J_{eff}=1/2$ limit, with octahedra that are nearly undistorted at low temperatures. Perhaps due to the smaller spin-orbit coupling expected in a $4d$ system, $\alpha$-RuCl$_3$ reveals an extremely narrow spin-orbit exciton (2 meV wide) well separated from charge excitations~\cite{Sandilands:2016gr}. Thus the low-energy model of $\alpha$-RuCl$_3$ does not contain any charge fluctuations, unlike the $5d$ Ir systems where the spin-orbit and onsite $d{-}d$ excitations are overlapped in energy~\cite{Gretarsson:2013fp}. Perhaps most promising is the observed continuum of magnetic excitations, where the Raman temperature dependence and the excitation dispersion seen by neutrons is consistent with fractional particles expected from the pure Kitaev model~\cite{Nasu:2016hga,Sandilands:2015hn,Glamazda:2017gc,Banerjee:2017dk}.

Despite its importance to the formation of an ordered state, the structure of $\alpha$-RuCl$_3$ remains controversial. In particular, the exact structure appears to be sensitive to atomic disorder and stacking faults, which are not uncommon in van der Waals crystals such as $\alpha$-RuCl$_3$~\cite{Cao:2016ep,Majumder:2015ck,Maksov:2016gr,Glamazda:2017gc}. Surprisingly, the addition of stacking faults leads to an enhanced onset of antiferromagnetic order (higher T$_N$)~\cite{Cao:2016ep}. This rather counterintuitive observation may result from additional tunneling pathways opened by the stacking disorder that boost the Heisenberg terms. If correct, this suggests that exfoliating $\alpha$-RuCl$_3$ down to single layers could suppress the long range order and perhaps stabilize the quantum spin liquid.

Through the use of applied magnetic fields, a number of probes now suggest that once the long range magnetic order is suppressed, a spin liquid state emerges~\cite{Leahy:2017cv,2017arXiv170607240P,2017arXiv170606157W}. If a similar suppression is achieved through exfoliation, the resulting state could be closer to the original Kitaev proposal. In addition, exfoliated $\alpha$-RuCl$_{3}$ would allow one to bring many well-developed techniques for exploring 2D atomic crystals to bear, for instance by utilizing van der Waals heterostructures to control the local environment and perhaps even exploit proximity effects~\cite{Geim:2013hf,Jariwala:2016el}. Crucial to this approach will be the evolution of the lattice and electronic structure that ultimately control the balance between the various exchange terms. Inelastic light scattering provides an excellent means to explore these issues. Raman has already demonstrated its utility in determining the lattice structure, symmetry, magnetic order, two-magnon excitations, magneto-elastic coupling, and disorder in various 2D materials~\cite{Tian:2016ki,Wang:2016rs,Sandilands:2010je,Lee:2016ga,Ferrari:2013jx}. In $\alpha$-RuCl$_3$, bulk measurements have revealed the spin-orbit exciton, and how magnetic fluctuations interact with the lattice via the asymmetric phonon broadening (i.e. Fano lineshapes)~\cite{Sandilands:2015hn,Glamazda:2017gc}. The Raman response also showed the hysteresis in the structural transition via the appearance/disappearance of modes as well as the mode frequency and strength~\cite{Glamazda:2017gc}. As with efforts to determine the structure via diffraction techniques, the polarization dependence of the modes has been interpreted in terms of both the single layer (undistorted honeycomb) and monoclinic (distorted) structures. As shown in Table~\ref{Tab:Select}, these two can be separated by the response of phonons to different polarization configurations of the incident and Raman scattered light. 

\begin{table}[t]
    \centering
    \begin{tabular}{c|c|c|c|c}
        Structure & Space Group & Possible Raman Modes & $c(XX)\bar{c}$ & $c(XY)\bar{c}$ \\\hline
         Trigonal & P31 & 23A + 23 E$_{1}$ +23 E$_{2}$ & A, E$_{1}$, E$_{2}$ & E$_{1}$, E$_{2}$ \\\hline
         Rhombohedral & R-3 & 4A$_{c}$ + 4E$_{g1}$ + 4E$_{g2}$ & A, E$_{g1}$, E$_{g2}$ & E$_{g1}$, E$_{g2}$ \\\hline
         Triclinic & P-1 & 12A$_{g}$ & A$_{g}$ & A$_{g}$ \\\hline
         Monoclinic & C2/M & 6A$_{g}$ + 6 B$_{g}$ & A$_{g}$ & A$_{g}$ \\\hline
    \end{tabular}
    \caption{\label{Tab:Select} Structures for $\alpha$-RuCl$_3$, along with the possible modes Raman allowed by symmetry. In addition we show the modes that would arise in back scattering for co-linear ($c(XX)\bar{c}$) and crossed ($c(XY)\bar{c}$) configurations. We note that only the Trigonal and Rhombohedral structures, with undistorted honeycombs, allow a mode to appear in the colinear---but not the crossed---polarization configurations.}
\end{table}

To directly address the possibility of tuning the behavior of $\alpha$-RuCl$_3$ via exfoliation, we performed a Raman study of several exfoliated crystals with thickness ranging from 43 nm down to single layers. We have measured the polarization and temperature dependence from room temperature to 10 K. While the thicker pieces produce spectra similar to those seen in bulk, in thin layers we observe an anomalous  broadening of the phonons with temperature. Particular attention is paid to a mode already shown to exhibit strong coupling to the magnetic excitations, suggesting an enhanced broadening from additional magnetic fluctuations in thin layers. Intriguingly, we also find an A$_{1g}$ phonon mode---forbidden by symmetry to appear in crossed polarization---is nonetheless observed in the thinnest samples at all temperatures. This is suggestive of a structural transition driven by the decreasing thickness or by strain incurred in the process of exfoliation.

\section{Exfoliation}
\begin{figure*}[t]
\begin{center}
\includegraphics[width=0.9\textwidth]{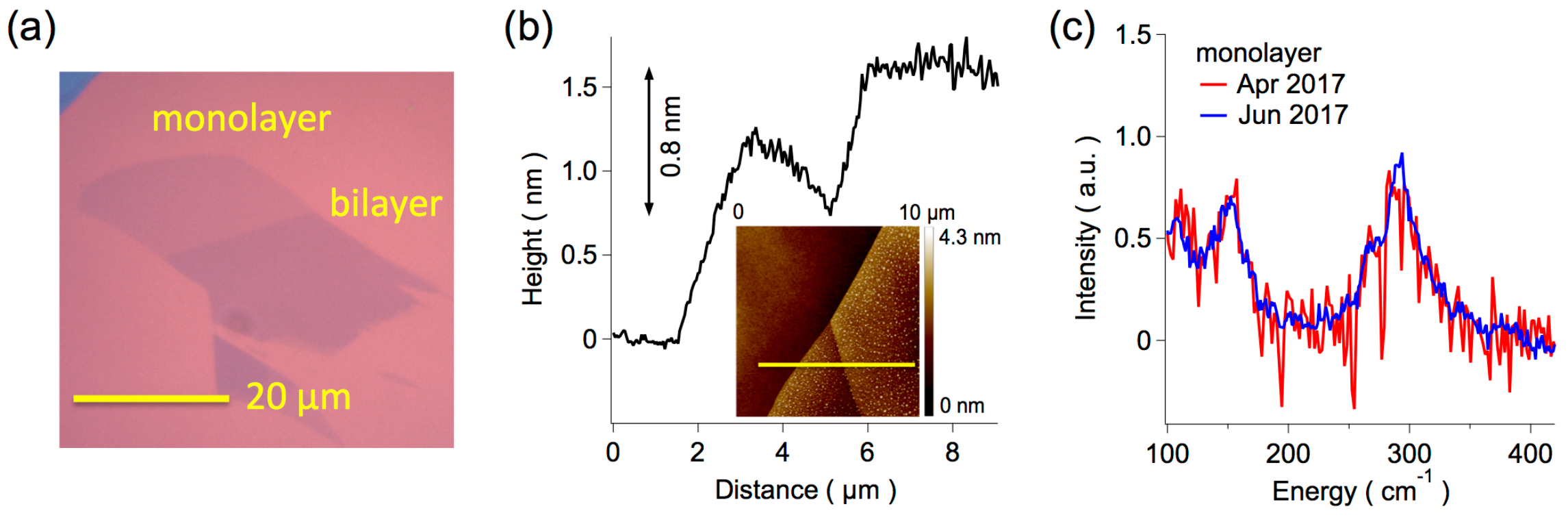}
\end{center}
\caption{\label{Exfoliation}Exfoliated $\alpha$-RuCl$_3$. (a) Optical image showing contrast of mono and bilayer flake. (b) Atomic force microscope image of same flake. (c) Air stability of monolayer $\alpha$-RuCl$_3$ as demonstrated by Raman spectroscopy: the red and blue traces were acquired in separate measurements made 2.5 months apart in different cryostats.}
\end{figure*}

Single crystals of $\alpha$-RuCl$_3$ were grown using a vapor transport technique from phase pure commercial RuCl$_3$ powder~\cite{Banerjee:2017dk}. Exfoliated crystals with typical sizes $10{-}30~\mu$m were mechanically exfoliated onto Si/SiO$_2$ substrates from bulk crystals using scotch tape following the methods used for graphene and other van der Waals materials~\cite{Novoselov:2005wx}. Exfoliation yielded monolayer, bilayer and multilayer flakes, which were first identified by optical color contrast (Figure~\ref{Exfoliation}a), and the thicknesses were then checked by atomic force microscopy (Figure~\ref{Exfoliation}b). Typical single layer thicknesses were 0.8 nm, compared to $\approx6$ \AA~found in xray diffraction~\cite{Cao:2016ep}. Fortuitously, unlike other recently explored vdW materials such as black phosphorous and MoS$_2$, $\alpha$-RuCl$_3$ appears to be stable in air: we find the Raman spectra of all thicknesses even down to monolayers to be reproducible after months of exposure to air, even when measured in separate Raman systems with different excitation lasers (see  Figure~\ref{Exfoliation}c).

\section{Raman}
Raman spectroscopy was performed on monolayer, bilayer and multilayer $\alpha$-RuCl$_3$ on Si/SiO2 at both room and low temperature. Room temperature spectra were measured in air in a Raman spectrometer with a 514 nm excitation laser and low energy cut-off of 100 cm$^{-1}$. The laser spot size was  $\approx2~\mu$m with a resolution of 1.8 cm$^{-1}$. Low temperature measurements were performed separately under vacuum in a cryostat down to 10 K. These measurements were made in the quasi-backscattering geometry in crossed (XY) polarization, with light polarized in the basal (cleavage) plane. Light from a 532 nm laser was focused down to a $2\ \mu$m spot with an estimated power of 160 ${\mu}$W. The resolution of the low temperature Raman system was 2 cm$^{-1}$. To confirm heating was not a contributing factor to the additional broadening, additional experiments were performed in a custom-built low temperature Raman setup where the anti-Stokes spectra could be collected~\cite{Tian:2016ja}. The ratio of the Stokes to anti-Stokes intensity confirmed the excitation power used did not lead to significant heating ($\Delta T\leq 10~K$). 

In measurements of multilayer stacks of thin-films, Fabry-Perot interference can result in strong deviations of the Raman intensity~\cite{1980PhRvL..44..273N,1989ApOpt..28.4017R,Blake:2007hb,Zhang:2015hv,Sandilands:2010je,Zhao:2011bs}, though we note these are unlikely to be strongly temperature dependent. Multiple reflections of both the incident excitation laser and of the Raman scattered light can occur due to the multiple interfaces present. If the optical constants of the constituent materials are known it is possible to compute the net effect this has on the spectra and then normalize it out. This computation consists of two main parts. First, the enhancements of the incident and scattered light at a given depth of the material are calculated. Then, these two enhancements are multiplied and integrated over the thickness of the sample to obtain the total enhancement factor. Each Raman spectrum is divided by the final enhancement factor, after a detector-dependent dark count was subtracted~\cite{Zhao:2011bs,Yoon:2009ke}.

Key to understanding the Raman results are their implications for the symmetry of the lattice structure. We use the standard Porto notation for back-scattering geometry $A(IS)\bar{A}$, where $A$ refers to the propagation axis of the light, $I$ is the axis along which the incident light is polarized, and $S$ is the polarization axis of the scattered light. From the different lattice structures we have calculated the possible symmetries of Raman active modes, and in which geometries they should be observed. These are shown in Table~\ref{Tab:Select}, but can be easily summarized by noting that the lattice structures fall into two categories: those without a distortion of the honeycomb lattice (Trigonal and Rhombohedral), versus those with an in plane distortion (Triclinic and Monoclinic). The undistorted lattices produce modes of A and E symmetry, with A modes only observed in collinear polarization (i.e. $c(XX)\bar{c}$), whereas the E modes are seen in both scattering geometries. For the distorted lattice structures, all observed modes should appear in both scattering configurations. Thus the original bulk data, and those reported here for thick exfoliated crystals wherein a mode is observed being suppressed in $c(XY)\bar{c}$, would be consistent with the undistorted in-plane lattice structures.

\section{Thickness and Temperature Dependence}
\begin{figure*}[t]
\begin{center}
\includegraphics[width=0.9\textwidth]{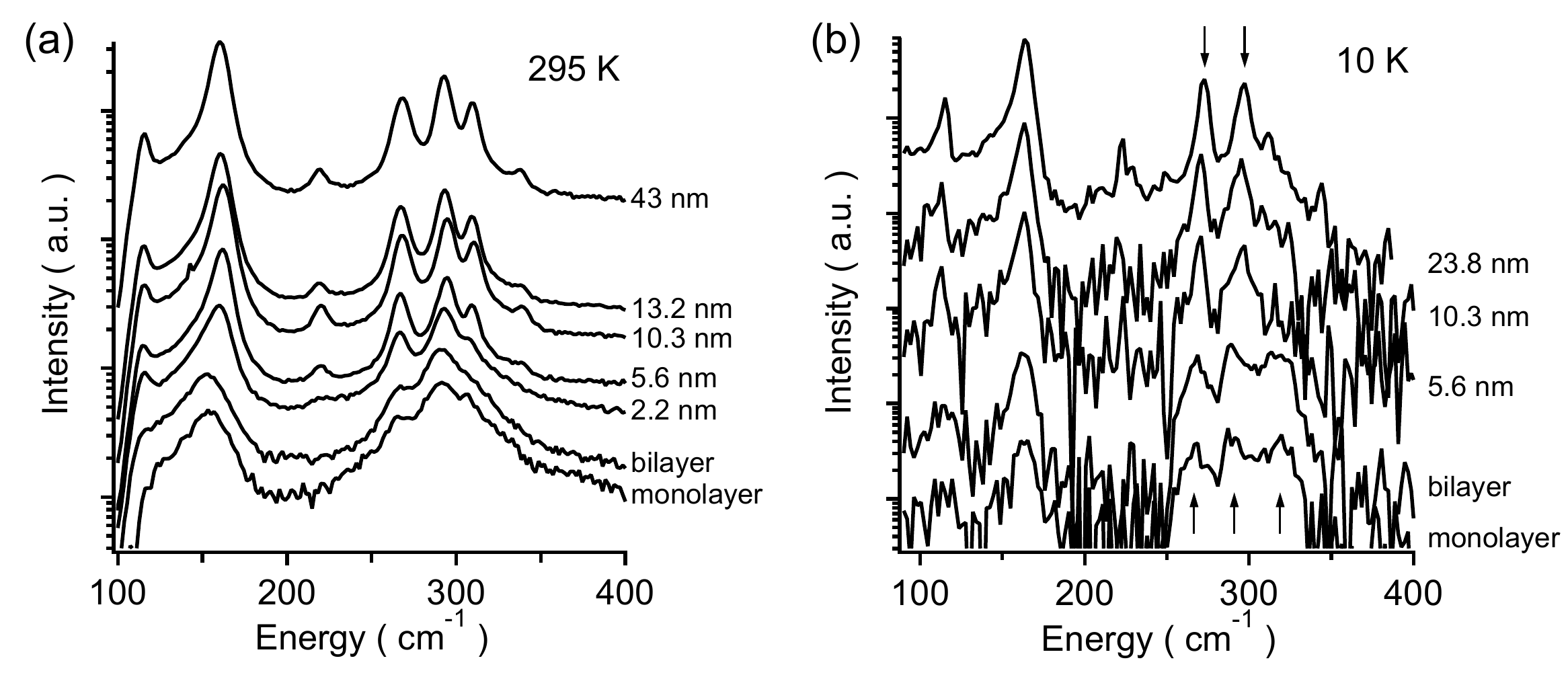}
\end{center}
\caption{\label{Thickness}Thickness dependence of Raman measurements: (a) unpolarized spectra at room temperature, and (b) in XY polarization at low temperature. Plots are shown as log intensity due to the large signal decrease with thickness. Traces in (b) are offset for clarity; no offsets are applied in (a).}
\end{figure*}

We begin with the thickness dependence of Raman spectroscopy of $\alpha$-RuCl$_3$ at high and low temperatures. In Figure~\ref{Thickness}a we show the thickness dependence of the unpolarized Raman spectra at room temperature, revealing both a nearly three order-of-magnitude decrease in the Raman response with reduced thickness, along with a notable increase in the broadening of all phonon modes for flakes thinner than $\approx6$ nm. As the broadening increases the three individual phonon peaks in the group between 270 and 315 cm$^{-1}$ merge. Fits to this data give the best results when three Lorentzians are used for all thicknesses, suggesting the individual modes still exist at room temperature. The source of this broadening is not immediately clear, but its occurrence in only the thinnest flakes suggests it may be related to the exfoliation process. For instance, prior work in graphene has demonstrated that exfoliated flakes show a range of residual strain~\cite{Bunch:2007ux} that may, in the present system, distort the lattice sufficiently to broaden the modes. In lieu of such global strain, the $\sim$nm-scale roughness of the supporting SiO$_2$ may induce a locally varying strain potential impacting thinner flakes more due to their reduced stiffness. Finally, the three weakest modes at 106 cm$^{-1}$, 205 cm$^{-1}$, and 315 cm$^{-1}$, appear to vanish in the thinnest flakes but may simply have become too broad to be distinguished from the noise. In short, beyond the loss of intensity and increased broadening, there is little change from the known spectra of bulk $\alpha$-RuCl$_3$~\cite{Sandilands:2015hn,Glamazda:2017gc}.

In contrast, the thickness dependence at low temperature shows an interesting anomaly (see Figure~\ref{Thickness}b). In XY polarization thicker flakes reveal only two peaks between 270 and 300 cm$^{-1}$ (marked by downward arrows), in agreement with results in bulk material~\cite{Sandilands:2015hn}. However, in the thinnest samples a third peak at 315 cm$^{-1}$ has appeared, in the same position as the third peak in the unpolarized and $c(XX)\bar{c}$ measurements of bulk samples. 

\begin{figure}[t]
    \centering
    \includegraphics[width=\textwidth]{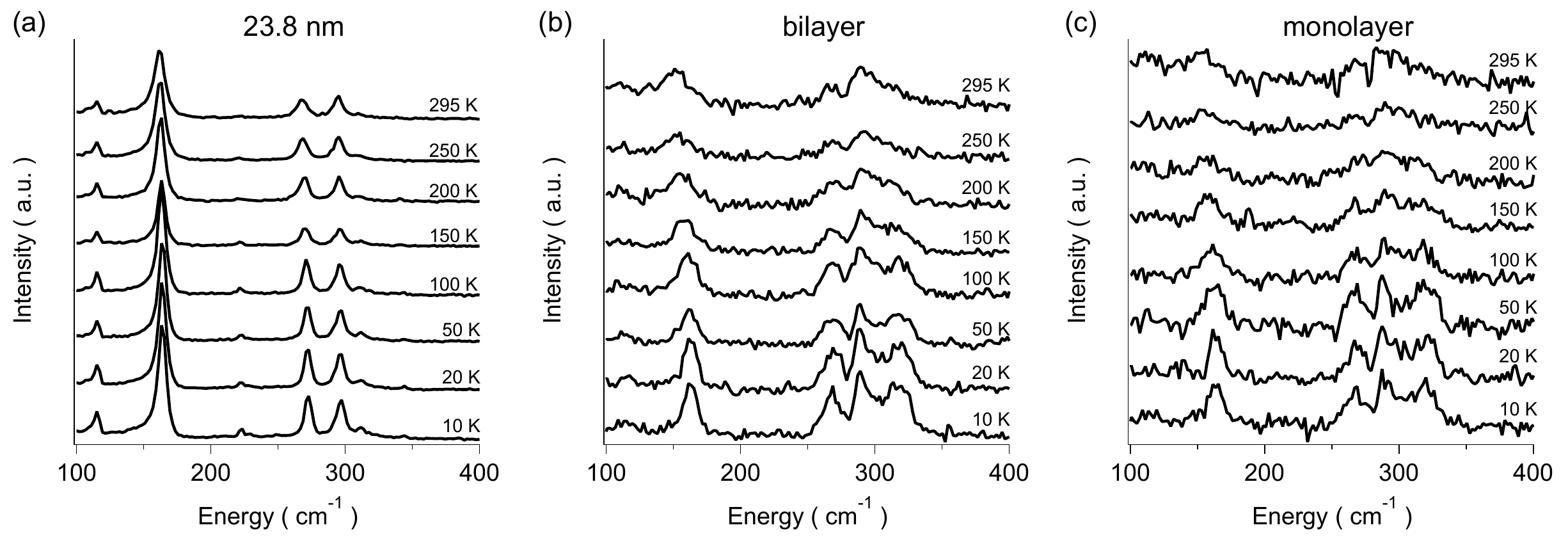}
    \caption{\label{Temperature}Temperature dependence of Raman spectra for various thickness flakes, measured in XY polarization: (a) bulk exfoliated crystal, 23.8 nm thick; (b) bilayer; and (c) monolayer.}
\end{figure}

To separate out static disorder from thermally induced dynamic fluctuations, we measured the temperature dependence of the Raman spectra in three samples: a `bulk' flake 23.8 nm thick, along with bilayer and monolayer flakes. These data, acquired in XY configuration, are shown in Figure~\ref{Temperature}, where we immediately observe a strong difference in the response of the bilayer and single layer from the thicker flake. Similar to previous bulk measurements performed in XY geometry, the 23.8 nm sample primarily reveals modes of E$_{g}$ symmetry  at all temperatures. The most pronounced effect seen occurs in the mode at 164 cm$^{-1}$, which acquires a significant Fano lineshape at low temperatures but does not reveal strong broadening with increased temperature. In contrast, in the thinnest layers (Figure~\ref{Temperature}b,c), a symmetry-forbidden mode (315 cm$^{-1}$) is clearly observed at low temperatures. This mode exhibits a curiously strong temperature dependence: while all phonons peaks are observed to sharpen with decreasing temperature, this 315 cm$^{-1}$ phonon is barely visible at all at room temperature, but rapidly grows in intensity as the temperature is lowered. The similar behavior of the mono- and bilayer flakes confirms this behavior is not an isolated event. Moreover, the mode at 164 cm$^{-1}$ in the thin samples undergoes a strong increase in width with temperature. 

\section{Discussion}
The appearance of the 315 cm$^{-1}$ mode in XY scattering is quite surprising. Only undistorted layers should contain both the E$_g$ and A$_{1g}$ modes, with the A$_{1g}$ mode disappearing in the XY configuration. Indeed, this was seen in Ref.~\cite{Sandilands:2015hn}, where a D3d symmetry of a single layer was assumed (which was also found to be consistent with the behavior observed for the magnetic continuum). In agreement with that prior work, in the thicker flake the two modes at 270 and 300 cm$^{-1}$ dominate in XY scattering. Interestingly, for distortions of a single layer consistent with the monoclinic structure, the high temperature bulk phase would produce Raman modes only of A$_{g}$ and B$_{g}$ symmetry; and in our geometry the A$_{g}$ modes should be visible in both polarization configurations. Noting that the thicker flake does show this nominally forbidden mode---albeit with a very low intensity---the observation of this mode may indicate the onset of a small distortion of the lattice at low temperatures, that is strongest in the thinnest flakes. In principle, this mode could also appear due to polarizer mis-alignment. However, in that case i) the observed strong temperature dependence would not be expected, and ii) the mode should appear in the thicker flake with more or less the same relative intensity as in the thinner flakes. 

\begin{figure}[t]
    \centering
    \includegraphics[width=0.8\textwidth]{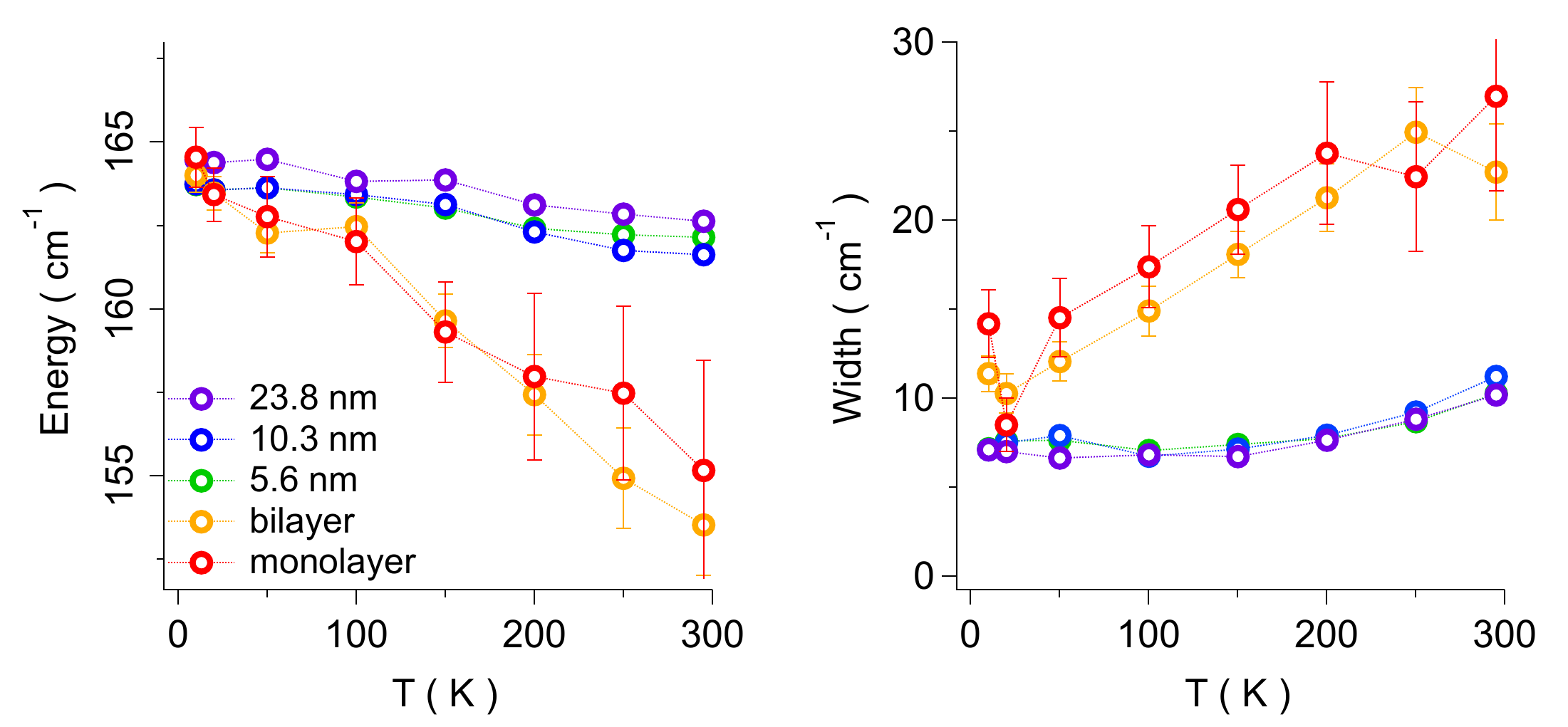}
    \caption{\label{Fano}Temperature dependence of the (a) energy and (b) width of the 164 cm$^{-1}$ phonon (labeled by its energy at low temperature), for five different flake thicknesses.}
\end{figure}

As noted, fits to these data strongly suggest the higher energy peak is present even at room temperature, although due to broadening it has merged with its neighbors. However the lower signal-to-noise in the thinner samples leads to large error bars in quantifying the intensity and width of the modes between 270 and 315 cm$^{-1}$. Therefore we turn our attention to the phonon mode at 164 cm$^{-1}$ visible in all samples, and which is established to strongly couple to the magnetic fluctuations~\cite{Sandilands:2015hn}.  In measurements of bulk $\alpha$-RuCl$_3$ crystals, this mode acquires a Fano lineshape ascribed to interactions with a continuum of magnetic excitations, and indeed a Fano lineshape is clearly visible in our data (see Figure~\ref{Temperature}). To quantify the change in the broadening of the mode with thickness, we performed fits to the data in a limited range around this peak. The results for the frequency and linewidth for five different thicknesses are shown in Figure~\ref{Fano}. For samples thicker than two layers we find behavior consistent with the bulk results, namely a small softening resulting from anharmonicity~\cite{1972PhRvB...6.1490L}, and a broadening above 100 K attributed to standard phonon decay into two acoustic modes~\cite{1966PhRv..148..845K}. The divergence of the mono- and bilayer samples' behavior from that of the thicker material is rather striking: both the central energies and widths are closely matched across all samples at low temperature. However, as the temperature increases the mono- and bilayer samples reveal a much stronger mode softening accompanied by a sharp increase of the width, which at room temperature is $2{-}3$ times broader than the same mode in thicker samples.

Thus overall we find the two thinnest samples stand out by the appearance of a symmetry-forbidden mode and an unusual behavior of the lower energy, Fano-shaped mode. Is a common cause behind these observations? Structural disorder is an obvious culprit: as mentioned, thin flakes exfoliated on Si/SiO$_2$ can have a range of built-in tension~\cite{Bunch:2007ux}, which here may introduce distortions of the Cl octahedra and enable otherwise forbidden phonons to appear. Indeed a monoclinic to rhombohedral structural transition has been observed in some bulk crystals below 150 K. These two structures lie close in energy and it is conceivable that the strain of exfoliation could re-introduce distortions tending toward the monoclinic structure~\cite{Cao:2016ep}. Interestingly, even at high temperatures, bulk-like samples provide little evidence for the mode in XY. Combined with our observation that at low temperatures the energies and linewidths of all modes are very close to that of the bulk, we rule out any major structural reorganization. Instead our results suggest  a stronger distortion appears only in thin layers. This agrees with prior x-ray and neutron scattering studies which find that even aggressive bending of bulk $\alpha$-RuCl$_3$ crystals results in only small structural changes~\cite{Cao:2016ep}.

On the other hand, the 164 cm$^{-1}$ phonon width may be broadened by a variety of physical mechanisms: disorder, an increase in anharmonic phonon scattering, and especially magnetic fluctuations which are already known to drive the Fano lineshape. Examining Figure~\ref{Fano}, we see at the lowest temperatures both the central energy and the linewidth values closely agree across all samples at the lowest temperatures. Since disorder will dominate as T$\to0$, it would appear that the disorder potential in the thinnest flakes is not particularly stronger than that affecting the thicker ones. Thus we may contemplate more exotic origins of the distortion, for instance, the possibility of a strong magneto-elastic coupling generating a distortion in order to stabilize magnetic order, driven by the large entropy of the proximate quantum spin liquid. Nonetheless, the strong temperature dependence of the linewidth with similar low temperature energy, suggests enhanced magnetic fluctuations cause strong additional broadening in thin layers. 

\section{Conclusion}
In this paper we explored the possibility of destroying the long range order in $\alpha$-RuCl$_3$ by exfoliating down to a single monolayer. The Raman spectra are consistent with the overall structure remaining intact, however the appearance of a mode in both XX and XY configurations suggest the development of significant in-plane distortions away from the perfect honeycomb lattice. If correct this suggests enhancement of non-Kitaev terms in the exchange that could further stabilize the ordered state. At the moment the origin of this distortion is unclear. The exfolation process, or the use of non-flat substrates, may contribute significant strain leading to the distorted structure, but further Raman and x-ray studies on various substrates~\cite{Lupascu:2012gm} will be required to clarify both the structure, and the role played by the substrate. We also note that the phonons in mono- and bilayer samples reveal significant additional temperature dependence. This could result from the additional distortion leading to enhanced anharmonicity. However a much more intriguing possibility is that enhanced magnetic fluctuations in thin layers strongly reduces the lifetime of the phonons. Studies on substrates with reduced background contributions and further enhancements of the Raman via interference would enable closer examination of the magnetic fluctuations. Nonetheless our results clearly demonstrate the possibility of tuning the Kitaev paramagnetic state in $\alpha$-RuCl$_3$ via exfoliation. 

\section{Acknowledgements}
Work at Boston College was completed with support from the National Science Foundation (grant DMR-1410846). BZ and EAH acknowledge support from the Institute of Materials Science \& Engineering at Washington University in St.~Louis. We are grateful to J. Knolle and N. Perkins for fruitful discussions. 
\section*{References}

\bibliographystyle{elsarticle-num}

\begin{thebibliography}{10}
\expandafter\ifx\csname url\endcsname\relax
  \def\url#1{\texttt{#1}}\fi
\expandafter\ifx\csname urlprefix\endcsname\relax\def\urlprefix{URL }\fi
\expandafter\ifx\csname href\endcsname\relax
  \def\href#1#2{#2} \def\path#1{#1}\fi

\bibitem{Kitaev:2006ik}
A.~Kitaev, {Anyons in an exactly solved model and beyond}, Annals of Physics
  321~(1) (2006) 2--111.

\bibitem{Nasu:2016hga}
J.~Nasu, J.~Knolle, D.~L. Kovrizhin, Y.~Motome, R.~Moessner, {Fermionic
  response from fractionalization in an insulating two-dimensional~magnet},
  Nature Physics 12~(10) (2016) 912--915.

\bibitem{Jackeli:2009hz}
G.~Jackeli, G.~Khaliullin, {Mott Insulators in the Strong Spin-Orbit Coupling
  Limit: From Heisenberg to a Quantum Compass and Kitaev Models}, Physical
  Review Letters 102~(1) (2009) 017205.

\bibitem{Kim:2008gi}
B.~J. Kim, H.~Jin, S.~J. Moon, J.~Y. Kim, B.~G. Park, C.~S. Leem, J.~Yu, T.~W.
  Noh, C.~Kim, S.~J. Oh, J.~H. Park, V.~Durairaj, G.~Cao, E.~Rotenberg, {Novel
  J$_{eff}=1/2$ Mott State Induced by Relativistic Spin-Orbit Coupling in
  Sr$_2$IrO$_4$}, Physical Review Letters 101~(7) (2008) 076402--4.

\bibitem{2017arXiv170107837C}
A.~Catuneanu, Y.~Yamaji, G.~Wachtel, H.-Y. Kee, Y.~B. Kim, {Realizing quantum
  spin liquid phases in spin-orbit driven correlated materials}, arXiv.org
  (2017) arXiv:1701.07837\href {http://arxiv.org/abs/1701.07837v1}
  {\path{arXiv:1701.07837v1}}.

\bibitem{Kim:2015iq}
H.-S. Kim, V.~S. V, A.~Catuneanu, H.-Y. Kee, {Kitaev magnetism in honeycomb
  RuCl$_3$ with intermediate spin-orbit coupling}, Physical Review B 91~(24)
  (2015) 241110.

\bibitem{Yadav:2016br}
R.~Yadav, N.~A. Bogdanov, V.~M. Katukuri, S.~Nishimoto, J.~van~den Brink,
  L.~Hozoi, {Kitaev exchange and field-induced quantum spin-liquid states in
  honeycomb $\alpha$-RuCl3}, Scientific Reports (2016) 1--16.

\bibitem{Mazin:2012gq}
I.~I. Mazin, H.~O. Jeschke, K.~Foyevtsova, R.~Valent{\'\i}, D.~I. Khomskii,
  {Na$_2$IrO$_3$ as a Molecular Orbital Crystal}, Physical Review Letters
  109~(19) (2012) 197201--5.

\bibitem{Gretarsson:2013fp}
H.~Gretarsson, J.~P. Clancy, X.~Liu, J.~P. Hill, E.~Bozin, Y.~Singh, S.~Manni,
  P.~Gegenwart, J.~Kim, A.~H. Said, D.~Casa, T.~Gog, M.~H. Upton, H.-S. Kim,
  J.~Yu, V.~M. Katukuri, L.~Hozoi, J.~van~den Brink, Y.-J. Kim, {Crystal-Field
  Splitting and Correlation Effect on the Electronic Structure of
  A$_2$IrO$_3$}, Physical Review Letters 110~(7) (2013) 076402--5.

\bibitem{Plumb:2014hh}
K.~W. Plumb, J.~P. Clancy, L.~J. Sandilands, V.~V. Shankar, Y.~F. Hu, K.~S.
  Burch, H.-Y. Kee, Y.-J. Kim, {$\alpha$-RuCl$_3$: A spin-orbit assisted Mott
  insulator on a honeycomb lattice}, Physical Review B 90~(4) (2014) 041112.

\bibitem{Sandilands:2015hn}
L.~J. Sandilands, Y.~Tian, K.~W. Plumb, Y.-J. Kim, K.~S. Burch, {Scattering
  Continuum and Possible Fractionalized Excitations in $\alpha$-RuCl$_3$},
  Physical Review Letters 114~(14) (2015) 147201.

\bibitem{Johnson:2015jc}
R.~D. Johnson, S.~C. Williams, A.~A. Haghighirad, J.~Singleton, V.~Zapf,
  P.~Manuel, I.~I. Mazin, Y.~Li, H.~O. Jeschke, R.~Valent, R.~Coldea,
  {Monoclinic crystal structure of $\alpha$-RuCl$_3$ and the zigzag
  antiferromagnetic ground state}, Physical Review B 92~(23) (2015) 1038--12.

\bibitem{Banerjee:2016jz}
A.~Banerjee, C.~A. Bridges, J.~Q. Yan, A.~A. Aczel, L.~Li, M.~B. Stone, G.~E.
  Granroth, M.~D. Lumsden, Y.~Yiu, J.~Knolle, S.~Bhattacharjee, D.~L.
  Kovrizhin, R.~Moessner, D.~A. Tennant, D.~G. Mandrus, S.~E. Nagler,
  {Proximate Kitaev quantum spin liquid behaviour in a honeycomb magnet},
  Nature Materials 15~(7) (2016) 733--740.

\bibitem{Sandilands:2016gr}
L.~J. Sandilands, Y.~Tian, A.~A. Reijnders, H.-S. Kim, K.~W. Plumb, Y.-J. Kim,
  H.-Y. Kee, K.~S. Burch, {Spin-orbit excitations and electronic structure of
  the putative Kitaev magnet $\alpha$-RuCl$_3$}, Physical Review B 93~(7)
  (2016) 075144.

\bibitem{Cao:2016ep}
H.~B. Cao, A.~Banerjee, J.~Q. Yan, C.~A. Bridges, M.~D. Lumsden, D.~G. Mandrus,
  D.~A. Tennant, B.~C. Chakoumakos, S.~E. Nagler, {Low-temperature crystal and
  magnetic structure of $\alpha$-RuCl$_3$}, Physical Review B 93~(13) (2016)
  134423.

\bibitem{Kubota:2015gu}
Y.~Kubota, H.~Tanaka, T.~Ono, Y.~Narumi, K.~Kindo, {Successive magnetic phase
  transitions in $\alpha$-RuCl$_3$: XY-like frustrated magnet on the honeycomb
  lattice}, Physical Review B 91~(9) (2015) 094422.

\bibitem{Lang:2016jn}
F.~Lang, P.~J. Baker, A.~A. Haghighirad, Y.~Li, D.~Prabhakaran,
  R.~Valent{\'\i}, S.~J. Blundell, {Unconventional magnetism on a honeycomb
  lattice in $\alpha$-RuCl$_3$ studied by muon spin rotation}, Physical Review
  B 94~(2) (2016) 020407.

\bibitem{Zhou:2016fe}
X.~Zhou, H.~Li, J.~A. Waugh, S.~Parham, H.-S. Kim, J.~A. Sears, A.~Gomes, H.-Y.
  Kee, Y.-J. Kim, D.~S. Dessau, {Angle-resolved photoemission study of the
  Kitaev candidate $\alpha$-RuCl$_3$}, Physical Review B 94~(16) (2016) 161106.

\bibitem{Ran:2017ke}
K.~Ran, J.~Wang, W.~Wang, Z.-Y. Dong, X.~Ren, S.~Bao, S.~Li, Z.~Ma, Y.~Gan,
  Y.~Zhang, J.~T. Park, G.~Deng, S.~Danilkin, S.-L. Yu, J.-X. Li, J.~Wen,
  {Spin-Wave Excitations Evidencing the Kitaev Interaction in Single
  Crystalline $\alpha$-RuCl$_3$}, Physical Review Letters 118 (2017) 107203.

\bibitem{Banerjee:2017dk}
A.~Banerjee, J.~Yan, J.~Knolle, C.~A. Bridges, M.~B. Stone, M.~D. Lumsden,
  D.~G. Mandrus, D.~A. Tennant, R.~Moessner, S.~E. Nagler, {Neutron scattering
  in the proximate quantum spin liquid $\alpha$-RuCl$_3$}, Science 356~(6342)
  (2017) 1055--1059.

\bibitem{Glamazda:2017gc}
A.~Glamazda, P.~Lemmens, S.~H. Do, Y.~S. Kwon, K.~Y. Choi, {Relation between
  Kitaev magnetism and structure in $\alpha$-RuCl$_3$}, Physical Review B 95
  (2017) 174429.

\bibitem{Majumder:2015ck}
M.~Majumder, M.~Schmidt, H.~Rosner, A.~A. Tsirlin, H.~Yasuoka, M.~Baenitz,
  {Anisotropic Ru$^{3+}$~$4d^5$ magnetism in the $\alpha$-RuCl$_3$ honeycomb
  system: Susceptibility, specific heat, and zero-field NMR}, Physical Review B
  91 (2015) 180401.

\bibitem{Maksov:2016gr}
A.~Maksov, T.~Berlijn, W.~Zhou, H.~B. Cao, J.~Q. Yan, C.~A. Bridges, D.~G.
  Mandrus, S.~E. Nagler, A.~P. Baddorf, M.~Ziatdinov, A.~Banerjee, S.~V.
  Kalinin, {Atomic-scale observation of structural and electronic orders in the
  layered compound $\alpha$-RuCl$_3$}, Nature Communications 7 (2016) 13774.

\bibitem{Leahy:2017cv}
I.~A. Leahy, C.~A. Pocs, P.~E. Siegfried, D.~Graf, S.~H. Do, K.-Y. Choi,
  B.~Normand, M.~Lee, {Anomalous Thermal Conductivity and Magnetic Torque
  Response in the Honeycomb Magnet $\alpha$-RuCl$_3$}, Physical Review Letters
  118 (2017) 187203.

\bibitem{2017arXiv170607240P}
A.~N. Ponomaryov, E.~Schulze, J.~Wosnitza, P.~Lampen-Kelley, A.~Banerjee, J.~Q.
  Yan, C.~A. Bridges, D.~G. Mandrus, S.~E. Nagler, A.~K. Kolezhuk, S.~A.
  Zvyagin, {Direct observation of the field-induced gap in the
  honeycomb-lattice material $\alpha$-RuCl$_3$}, arXiv.org (2017)
  arXiv:1706.07240\href {http://arxiv.org/abs/1706.07240}
  {\path{arXiv:1706.07240}}.

\bibitem{2017arXiv170606157W}
Z.~Wang, S.~Reschke, D.~H{\"u}vonen, S.~H. Do, K.~Y. Choi, M.~Gensch, U.~Nage,
  T.~R{\~o}{\~o}m, A.~Loidl, {Magnetic Excitations and Continuum of a
  Field-Induced Quantum Spin Liquid in $\alpha$-RuCl$_3$}, arXiv.org (2017)
  arXiv:1706.06157\href {http://arxiv.org/abs/1706.06157}
  {\path{arXiv:1706.06157}}.

\bibitem{Geim:2013hf}
A.~K. Geim, I.~V. Grigorieva, {Van der Waals heterostructures}, Nature
  499~(7459) (2013) 419--425.

\bibitem{Jariwala:2016el}
D.~Jariwala, T.~J. Marks, M.~C. Hersam, {Mixed-dimensional van der Waals
  heterostructures}, Nature Materials 16~(2) (2016) 170--181.

\bibitem{Tian:2016ki}
Y.~Tian, M.~J. Gray, H.~Ji, R.~J. Cava, K.~S. Burch, {Magneto-elastic coupling
  in a potential ferromagnetic 2D atomic crystal}, 2D Materials 3 (2017)
  025035.

\bibitem{Wang:2016rs}
X.~Wang, K.~Du, Y.~Y.~F. Liu, P.~Hu, J.~Zhang, Q.~Zhang, M.~H.~S. Owen, X.~Lu,
  C.~K. Gan, P.~Sengupta, C.~Kloc, Q.~Xiong, {Raman spectroscopy of atomically
  thin two-dimensional magnetic iron phosphorus trisulfide (FePS$_3$)
  crystals}, 2D Materials 3 (2016) 031009.

\bibitem{Sandilands:2010je}
L.~J. Sandilands, J.~X. Shen, G.~M. Chugunov, S.~Y.~F. Zhao, S.~Ono, Y.~Ando,
  K.~S. Burch, {Stability of exfoliated
  Bi$_2$Sr$_2$Dy$_x$Ca$_{1-x}$Cu$_2$O$_{8+\delta}$ studied by Raman
  microscopy}, Physical Review B 82 (2010) 064503.

\bibitem{Lee:2016ga}
J.-U. Lee, S.~Lee, J.~H. Ryoo, S.~Kang, T.~Y. Kim, P.~Kim, C.-H. Park, J.-G.
  Park, H.~Cheong, {Ising-Type Magnetic Ordering in Atomically Thin FePS$_3$},
  Nano Letters 16~(12) (2016) 7433.

\bibitem{Ferrari:2013jx}
A.~C. Ferrari, D.~M. Basko, {Raman spectroscopy as a versatile tool for
  studying the properties of graphene}, Nature Nanotechnology 8~(4) (2013)
  235--246.

\bibitem{Novoselov:2005wx}
K.~S. Novoselov, D.~Jiang, F.~Schedin, T.~J. Booth, V.~V. Khotkevich, S.~V.
  Morozov, A.~K. Geim, {Two-dimensional atomic crystals}, Proceedings of the
  National Academy of Sciences 102~(30) (2005) 10451.

\bibitem{Tian:2016ja}
Y.~Tian, A.~A. Reijnders, G.~B. Osterhoudt, I.~Valmianski, J.~G. Ramirez,
  C.~Urban, R.~Zhong, J.~Schneeloch, G.~Gu, I.~Henslee, K.~S. Burch, {Low
  vibration high numerical aperture automated variable temperature Raman
  microscope}, Review of Scientific Instruments 87~(4) (2016) 043105.

\bibitem{1980PhRvL..44..273N}
R.~J. Nemanich, C.~C. Tsai, G.~A.~N. Connell, {Interference-enhanced Raman
  scattering of very thin titanium and titanium oxide films}, Physical Review
  Letters 44~(4) (1980) 273--276.

\bibitem{1989ApOpt..28.4017R}
M.~Ramsteiner, C.~Wild, J.~Wagner, {Interference effects in the Raman
  scattering intensity from thin films}, Applied Optics 28~(18) (1989)
  4017--4023.

\bibitem{Blake:2007hb}
P.~Blake, A.~H. Castro~Neto, K.~S. Novoselov, D.~Jiang, R.~Yang, A.~K. Geim,
  {Making graphene visible}, Applied Physics Letters 91~(6) (2007) 063124.

\bibitem{Zhang:2015hv}
H.~Zhang, Y.~Wan, Y.~Ma, W.~Wang, Y.~Wang, L.~Dai, {Interference effect on
  optical signals of monolayer MoS$_2$}, Applied Physics Letters 107~(10)
  (2015) 101904--4.

\bibitem{Zhao:2011bs}
S.~Y.~F. Zhao, C.~Beekman, L.~J. Sandilands, J.~E.~J. Bashucky, D.~Kwok,
  N.~Lee, A.~D. LaForge, S.~W. Cheong, K.~S. Burch, {Fabrication and
  characterization of topological insulator Bi$_2$Se$_3$ nanocrystals}, Applied
  Physics Letters 98~(14) (2011) 141911.

\bibitem{Yoon:2009ke}
D.~Yoon, H.~Moon, Y.-W. Son, J.~S. Choi, B.~H. Park, Y.~H. Cha, Y.~D. Kim,
  H.~Cheong, {Interference effect on Raman spectrum of graphene on SiO$_2$/Si},
  Physical Review B 80~(12) (2009) 125422.

\bibitem{Bunch:2007ux}
J.~S. Bunch, A.~M. Van Der~Zande, S.~Verbridge, I.~Frank, D.~Tanenbaum, J.~M.
  Parpia, H.~G. Craighead, P.~L. McEuen, {Electromechanical resonators from
  graphene sheets}, Science 315~(5811) (2007) 490.

\bibitem{1972PhRvB...6.1490L}
R.~P. Lowndes, {Anharmonicity in the Silver and Thallium Halides: Far-Infrared
  Dielectric Response}, Physical Review B 6~(4) (1972) 1490--1498.

\bibitem{1966PhRv..148..845K}
P.~G. Klemens, {Anharmonic Decay of Optical Phonons}, Physical Review 148~(2)
  (1966) 845--848.

\bibitem{Lupascu:2012gm}
A.~Lupascu, R.~Feng, L.~J. Sandilands, Z.~Nie, V.~Baydina, G.~Gu, S.~Ono,
  Y.~Ando, D.~C. Kwok, N.~Lee, S.~W. Cheong, K.~S. Burch, Y.-J. Kim,
  {Structural study of Bi$_2$Sr$_2$CaCu$_2$O$_{8+\delta}$ exfoliated
  nanocrystals}, Applied Physics Letters 101~(22) (2012) 223106.

\end{thebibliography}

\end{document}